\documentclass[12pt,twoside]{article}
\usepackage{amssymb}
\usepackage{amsmath}
\usepackage{theorem}
\def\1{{1\mskip-10mu1}}

\def\bea{\begin{eqnarray*}}
\def\eea{\end{eqnarray*}}
\def\bean{\begin{eqnarray}}
\def\eean{\end{eqnarray}}

\newtheorem{ftheo}{THEOREM}[section]

\newtheorem{flemma}[ftheo]{LEMMA}

\newtheorem{frem}[ftheo]{REMARK}
\begin{document}

\author{Karl-Heinz Fichtner \\
Friedrich-Schiller-Universit\"{a}t Jena \\
Fakult\"{a}t f\"{u}r Mathematik und Informatik \\
Institut f\"{u}r Angewandte Mathematik \\
D-07740 Jena Deutschland \\
E-mail: fichtner@minet.uni-jena.de\\
and\\
Masanori Ohya\\
Department of Information Sciences\\
Science University of Tokyo\\
Chiba 278-8510 Japan\\
E-mail: ohya@is.noda.sut.ac.jp}
\title{Quantum Teleportation and Beam Splitting}
\date{}
\maketitle

\begin{abstract}
Following the previous paper in which quantum teleportation is rigorously
discussed with coherent entangled states given by beam splittings, we
further discuss two types of models, perfect teleportation model and
non-perfect teleportation model, in general scheme. Then the difference
among several models, i.e., the perfect models and the non-perfect models,
is studied. Our teleportation models are constructed by means of coherent
states in some Fock space with counting measures, so that our model can be
treated in the frame of usual optical communication.
\end{abstract}

\section{Introduction}

Following the previous paper \cite{FO}, we further discuss the non-perfect
teleportation. The notion of non-perfect teleportation is introduced in \cite
{FO} to construct a handy (i.e., physically more realizable) teleportation,
although its mathematics becomes a little more complicated. For the
completeness of the present paper, we quickly review the meaning of the
teleportation and some basic facts of Fock space in this section. Then we
dicuss the perfect teleportation in very general (more general than one
given in \cite{FO}) scheme with our previous results, and we state the main
theorem obtained in \cite{FO} for non-perfect teleportation, both in the
section 2. The main results of this paper are presented in the section 3,
where we discuss the difference among three models, i.e., the perfect model,
the non-perfect one given in \cite{FO} and that discussed in the present
paper. The proofs of the main results are given in the section 4.

\subsection{Quantum teleportation}

The study of quantum teleportation was started by the paper \cite{Ben} as a
part of quantum cryptography \cite{Eke}, whose scheme can be mathematically
expressed in the following steps \cite{IOS,AO2,FO}:

\begin{description}
\item[Step 0:]  A girl named Alice has an unknown quantum state $\rho $ on
(a $N$--dimensional) Hilbert space $\mathcal{H}_{1}$ and she was asked to
teleport it to a boy named Bob.

\item[Step 1:]  For this purpose, we need two other Hilbert spaces $\mathcal{%
H}_{2}$ and $\mathcal{H}_{3}$, $\mathcal{H}_{2}$ is attached to Alice and $%
\mathcal{H}_{3}$ is attached to Bob. Prearrange a so-called entangled state $%
\sigma $ on $\mathcal{H}_{2}\otimes \mathcal{H}_{3}$ having certain
correlations and prepare an ensemble of the combined system in the state $%
\rho \otimes \sigma $ on $\mathcal{H}_{1}\otimes \mathcal{H}_{2}\otimes 
\mathcal{H}_{3}$.

\item[Step 2:]  One then fixes a family of mutually orthogonal projections $%
(F_{nm})_{n,m=1}^{N}$ on the Hilbert space $\mathcal{H}_{1}\otimes \mathcal{H%
}_{2}$ corresponding to an observable $F:=\sum\limits_{n,m}z_{n,m}F_{nm}$,
and for a fixed one pair of indices $n,m$, Alice performs a first kind
incomplete measurement, involving only the $\mathcal{H}_{1}\otimes \mathcal{H%
}_{2}$ part of the system in the state $\rho \otimes \sigma $, which filters
the value $z_{nm}$, that is, after measurement on the given ensemble $\rho
\otimes \sigma $ of identically prepared systems, only those where $F$ shows
the value $z_{nm}$ are allowed to pass. According to the von Neumann rule,
after Alice's measurement, the state becomes 
\[
\rho _{nm}^{(123)}:=\frac{(F_{nm}\otimes \mathbf{1})\rho \otimes \sigma
(F_{nm}\otimes \mathbf{1})}{\mathrm{tr}_{123}(F_{nm}\otimes \mathbf{1})\rho
\otimes \sigma (F_{nm}\otimes \mathbf{1})} 
\]
where $\mathrm{tr}_{123}$ is the full trace on the Hilbert space $\mathcal{H}%
_{1}\otimes \mathcal{H}_{2}\otimes \mathcal{H}_{3}$.

\item[Step 3:]  Bob is informed which measurement was done by Alice. This is
equivalent to transmit the information that the eigenvalue $z_{nm}$ was
detected. This information is transmitted from Alice to Bob without
disturbance and by means of classical tools.

\item[Step 4:]  Making only partial measurements on the third part on the
system in the state $\rho _{nm}^{(123)}$ means that Bob will control a state 
$\Lambda _{nm}(\rho )$ on $\mathcal{H}_{3}$ given by the partial trace on $%
\mathcal{H}_{1}\otimes \mathcal{H}_{2}$ of the state $\rho _{nm}^{(123)}$
(after Alice's measurement) 
\begin{eqnarray*}
\Lambda _{nm}(\rho ) &=&\mathrm{tr}_{12}\;\rho _{nm}^{(123)} \\
&=&\mathrm{tr}_{12}\frac{(F_{nm}\otimes \mathbf{1})\rho \otimes \sigma
(F_{nm}\otimes \mathbf{1})}{\mathrm{tr}_{123}(F_{nm}\otimes \mathbf{1})\rho
\otimes \sigma (F_{nm}\otimes \mathbf{1)}}
\end{eqnarray*}
Thus the whole teleportation scheme given by the family $(F_{nm})$ and the
entangled state $\sigma $ can be characterized by the family $(\Lambda
_{nm}) $ of channels from the set of states on $\mathcal{H}_{1}$ into the
set of states on $\mathcal{H}_{3}$ and the family $(p_{nm})$ given by 
\[
p_{nm}(\rho ):=\mathrm{tr}_{123}(F_{nm}\otimes \mathbf{1})\rho \otimes
\sigma (F_{nm}\otimes \mathbf{1}) 
\]
of the probabilities that Alice's measurement according to the observable $F$
will show the value $z_{nm}$.
\end{description}

The teleportation scheme works perfectly with respect to a certain class $%
\frak{S}$ of states $\rho $ on $\mathcal{H}_{1}$ if the following conditions
are fulfilled.

\begin{description}
\item[(E1)]  For each $n,m$ there exists a unitary operator $v_{nm}:\mathcal{%
H}_{1}\to \mathcal{H}_{3}$ such that 
\[
\Lambda _{nm}(\rho )=v_{nm}\;\rho \;v_{nm}^{*}\quad (\rho \in \frak{S}) 
\]

\item[(E2)]  
\[
\sum\limits_{nm}p_{nm}(\rho )=1\quad (\rho \in \frak{S}) 
\]

\item  (E1) means that Bob can reconstruct the original state $\rho $ by
unitary keys $\{v_{nm}\}$ provided to him. \newline
\end{description}

(E2) means that Bob will succeed to find a proper key with certainty. 
\newline
Such a teleportation process can be classified into two cases \cite{AO2},
i.e., weak teleportation and general teleportation, in which the solutions
of the teleportation in each case and the conditions of the uniqueness of
unitary key were discussed. The solution of the weak teleportation is a
triple $\left\{ \sigma ^{(23)},F^{(12)},U\right\} $such that 
\[
\Lambda ^{*}\rho ^{(1)}=U^{*}\rho ^{(1)}U 
\]
holds for any state $\rho ^{(1)}\in \mathcal{S}(\mathcal{H}_{1})$ . Once a
weak solution of a teleportation problem is given, we can construct the
general solution for all $n,m$ above \cite{AO2}.

In \cite{FO}, we considered a teleportation model where the entangled state $%
\sigma $ is given by the splitting of a superposition of certain coherent
states, although this model doesn't work perfectly, that is, neither (E2)
nor (E1) hold. In the same paper, we estimated the difference between the
perfect teleportation and this non-perfect teleportation by adding a further
step in the teleportation scheme:

\begin{description}
\item[Step 5:]  Bob will perform a measurement on his part of the system
according to the projection 
\[
F_{+}:=\mathbf{1}-|\mathrm{exp}(0)><\mathrm{exp}(0)| 
\]
where $|\mathrm{exp}(0)><\mathrm{exp}(0)|$ denotes the vacuum state (the
coherent state with density $0$).
\end{description}

Then our new teleportation channels (we denote it again by $\Lambda _{nm}$)
have the form 
\[
\Lambda _{nm}(\rho ):=\mathrm{tr}_{12}\frac{(F_{nm}\otimes F_{+})\rho
\otimes \sigma (F_{nm}\otimes F_{+})}{\mathrm{tr}_{123}(F_{nm}\otimes
F_{+})\rho \otimes \sigma (F_{nm}\otimes F_{+})} 
\]
and the corresponding probabilities are 
\[
p_{nm}(\rho ):=\mathrm{tr}_{123}(F_{nm}\otimes F_{+})\,\rho \otimes \sigma
(F_{nm}\otimes F_{+}) 
\]
For this teleportation scheme, (E1) is fulfilled but (E2) is not, about
which we review in the next section.

\subsection{Basic Notions and Notations}

We collect some basic facts concerning the (symmetric) Fock space in a way
adapted to the language of counting measures. For details we refer to \cite
{FF1,FF2,FFL,AO2,L}. \newline

Let $G$ be an arbitrary complete separable metric space. Further, let $\mu $
be a locally finite diffuse measure on $G$, i.e. $\mu (B)<+\infty $ for
bounded measurable subsets of $G$ and $\mu (\{x\})=0$ for all singletons $%
x\in G$.

We denote the set of all finite counting measures on $G$ by $M=M(G)$. Since $%
\varphi \in M$ can be written in the form $\varphi
=\sum\limits_{j=1}^{n}\delta _{x_{j}}$ for some $n=0,1,2,\ldots $ and $%
x_{j}\in G$ with the Dirac measure $\delta _{x}$ corresponding to $x\in G$,
the elements of $M$ can be interpreted as finite (symmetric) point
configurations in $G$. We equip $M$ with its canonical $\sigma $--algebra $%
\frak{W}$ (cf. \cite{FF1}, \cite{FF2}) and we consider the $\sigma $--finite
measure $F$ by setting 
\[
F(Y):=\mathcal{X}_{Y}(O)+\sum\limits_{n\ge 1}\frac{1}{n!}\int\limits_{G^{n}}%
\mathcal{X}_{Y}\left( \sum\limits_{j=1}^{n}\delta _{x_{j}}\right) \mu
^{n}(d[x_{1},\ldots ,x_{n}])(Y\in \frak{W}), 
\]
where $\mathcal{X}_{Y}$ denotes the indicator function of a set $Y$ and $O$
represents the empty configuration, i.~e., $O(G)=0$.

Since $\mu $ was assumed to be diffuse one easily checks that $F$ is
concentrated on the set of a simple configurations (i.e., without multiple
points) 
\[
\hat{M}:=\{\varphi \in M|\varphi (\{x\})\le 1\text{ for all }x\in G\} 
\]

$\mathcal{M}=\mathcal{M}(G):=L^{2}(M,\frak{W},F)$ is called the (symmetric)
Fock space over $G$.

In \cite{FF1} it was proved that $\mathcal{M}$ and the Boson Fock space $%
\Gamma (L^{2}(G))$ in the usual definition are isomorphic. For each $\Phi
\in \mathcal{M}$ with $\Phi \neq 0$ we denote by $|\Phi >$ the corresponding
normalized vector 
\[
|\Phi >:=\frac{\Phi }{||\Phi ||}. 
\]
Further, $|\Phi ><\Phi |$ denotes the corresponding one--dimensional
projection describing a pure state given by the normalized vector $|\Phi >$.
Now, for each $n\ge 1$ let $\mathcal{M}^{\otimes n}$ be the $n$--fold tensor
product of the Hilbert space $\mathcal{M}$, which can be identified with $%
L^{2}(M^{n},F^{n})$.

\label{def2} For a given function $g:G\to \Bbb{C}$ the function $\mathrm{exp}%
\;(g):M\to \Bbb{C}$ defined by 
\[
\mathrm{exp}\;(g)\,(\varphi ):=\left\{ 
\begin{array}{lll}
1 & \text{ if } & \varphi =0 \\ 
\prod_{x\in G,\varphi \left( \left\{ x\right\} \right) >0}g(x) &  & otherwise
\end{array}
\right. 
\]
is called exponential vector generated by $g$.

Observe that $\mathrm{exp}\;(g)\in \mathcal{M}$ if and only if $g\in
L^{2}(G) $ and one has in this case \newline
$||\mathrm{exp}\;(g)||^{2}=e^{\Vert g\Vert ^{2}}$ and $|\mathrm{exp}
\;(g)>=e^{-\frac{1}{2}\Vert g\Vert ^{2}}\mathrm{exp}\;(g)$. The projection
\noindent \noindent $|\mathrm{exp}\;(g)><\mathrm{exp}\;(g)|$ is called the
coherent state corresponding to $g\in L^{2}(G)$. In the special case $%
g\equiv 0$ we get the vacuum state 
\[
|\mathrm{exp}(0)>=\mathcal{X}_{\{0\}}\;. 
\]
The linear span of the exponential vectors of $\mathcal{M}$ is dense in $%
\mathcal{M}$, so that bounded operators and certain unbounded operators can
be characterized by their actions on exponential vectors.

\label{def3} The operator $D:\mathrm{dom}(D)\to \mathcal{M}^{\otimes 2}$
given on a dense domain $\mathrm{dom}(D)\subset \mathcal{M}$ containing the
exponential vectors from $\mathcal{M}$ by 
\[
D\psi (\varphi _{1},\varphi _{2}):=\psi (\varphi _{1}+\varphi _{2})\quad
(\psi \in \mathrm{dom}(D),\,\varphi _{1},\varphi _{2}\in M) 
\]
is called compound Malliavin derivative. On exponential vectors $\mathrm{exp}%
\;(g)$ with $g\in L^{2}(G),$ one gets immediately 
\begin{equation}
D\;\mathrm{exp}\;(g)=\mathrm{exp}\;(g)\otimes \;\mathrm{exp}\;(g)  \label{1}
\end{equation}

The operator $S:\mathrm{dom}(S)\to \mathcal{M}$ given on a dense domain $%
\mathrm{dom}\;(S)\subset \mathcal{M}^{\otimes 2}$ containing tensor products
of exponential vectors by 
\[
S\Phi (\varphi ):=\sum\limits_{\tilde{\varphi}\le \varphi }\Phi (\tilde{%
\varphi},\varphi -\tilde{\varphi})\quad (\Phi \in \mathrm{dom}(S),\;\varphi
\in M) 
\]
is called compound Skorohod integral. One gets 
\begin{equation}
\langle D\psi ,\Phi \rangle _{\mathcal{M}^{\otimes 2}}=\langle \psi ,S\Phi
\rangle _{\mathcal{M}}\quad (\psi \in \mathrm{dom}(D),\;\Phi \in \mathrm{dom}%
(S))  \label{2}
\end{equation}
\begin{equation}
S(\mathrm{exp}\;(g)\otimes \mathrm{exp}\;(h))=\mathrm{exp}\;(g+h)\quad
(g,h\in L^{2}(G))  \label{3}
\end{equation}
For more details we refer to \cite{FW}.

\label{def5} Let $T$ be a linear operator on $L^{2}(G)$ with $\Vert T\Vert
\le 1$. Then the operator $\Gamma (T)$ called second quantization of $T$ is
the (uniquely determined) bounded operator on $\mathcal{M}$ fulfilling 
\[
\Gamma (T)\mathrm{exp}\;(g)=\mathrm{exp}\;(Tg)\quad (g\in L^{2}(G)). 
\]
Clearly, it holds 
\begin{eqnarray}
\Gamma (T_{1})\Gamma (T_{2}) &=&\Gamma (T_{1}T_{2})  \label{4} \\
\Gamma (T^{*}) &=&\Gamma (T)^{*}  \nonumber
\end{eqnarray}
It follows that $\Gamma (T)$ is an unitary operator on $\mathcal{M}$ if $T$
is an unitary operator on $L^{2}(G)$.

In \cite{FO} we proved.

\begin{flemma}
\label{def6} Let $K_{1},K_{2}$ be linear operators on $L^{2}(G)$ with
property 
\begin{equation}
K_{1}^{*}K_{1}+K_{2}^{*}K_{2}=\mathbf{1}\;.  \label{5}
\end{equation}
Then there exists exactly one isometry $\nu _{K_{1},K_{2}}$ from $\mathcal{M}
$ to $\mathcal{M}^{\otimes 2}=\mathcal{M}\otimes \mathcal{M}$ with 
\begin{equation}
\nu _{K_{1},K_{2}}\mathrm{exp}\;(g)=\mathrm{exp}(K_{1}g)\otimes \mathrm{exp}%
(K_{2}g)\quad (g\in L^{2}(G))  \label{6}
\end{equation}
Further it holds 
\begin{equation}
\nu _{K_{1},K_{2}}=(\Gamma (K_{1})\otimes \Gamma (K_{2}))D  \label{7}
\end{equation}
(at least on $\mathrm{dom}(D)$ but one has the unique extension). \newline
The adjoint $\nu _{K_{1},K_{2}}^{*}$ of $\nu _{K_{1},K_{2}}$ is
characterized by 
\begin{equation}
\nu _{K_{1},K_{2}}^{*}(\mathrm{exp}\;(h)\otimes \mathrm{exp}\;(g))=\mathrm{\
exp}(K_{1}^{*}h+K_{2}^{*}g)\quad (g,h\in L^{2}(G))  \label{8}
\end{equation}
and it holds 
\begin{equation}
\nu _{K_{1},K_{2}}^{*}=S(\Gamma (K_{1}^{*})\otimes \Gamma (K_{2}^{*}))
\label{9}
\end{equation}
\end{flemma}

From $K_{1},K_{2}$ we get a transition expectation $\xi _{K_{1}K_{2}}:%
\mathcal{M}\otimes \mathcal{M}\to \mathcal{M}$, using $\nu _{K_{1},K_{2}}$
and the lifting $\xi _{K_{1}K_{2}}^{*}$ may be interpreted as a certain
splitting (cf. \cite{AO2}). The property (\ref{5}) implies 
\begin{equation}
\Vert K_{1}g\Vert ^{2}+\Vert K_{2}g\Vert ^{2}=\Vert g\Vert ^{2}\quad (g\in
L^{2}(G))
\end{equation}
Let $U$, $V$ be unitary operators on $L^{2}(G)$. If operators $K_{1},K_{2}$
satisfy (\ref{5}),~then the pair $\hat{K}_{1}=UK_{1},\ \hat{K}_{2}=VK_{2}$
fulfill (\ref{5}).

Here we explain fundamental scheme of beam splitting \cite{FFL}. We define
an isometric operator $V_{\alpha ,\beta }$ for coherent vectors such that 
\[
V_{\alpha ,\beta }|\,\mathrm{exp}\;(g)\rangle =|\,\mathrm{exp}\;(\alpha
g)\rangle \otimes |\,\mathrm{exp}\,(\beta g)\rangle 
\]
with $\mid \alpha \mid ^{2}+\mid \beta \mid ^{2}=1$. This beam splitting is
a useful mathematical expression for optical communication and quantum
measurements \cite{AO2}. As one example, take $\alpha =\beta =\frac{1}{2}$
in the above formula and let $K_{1}=K_{2}$ be the following operator of
multiplication on $L^{2}(G)$ 
\[
K_{1}g=\frac{1}{\sqrt{2}}\;g=K_{2}g\quad (g\in L^{2}(G)) 
\]
We put 
\[
\nu :=\nu _{K_{1},K_{2}}, 
\]
then we obtain 
\begin{equation}
\nu \;\mathrm{exp}\;(g)=\mathrm{exp}\;\left( \frac{1}{\sqrt{2}}g\right)
\otimes \mathrm{exp}\;(\frac{1}{\sqrt{2}}\;g)\quad (g\in L^{2}(G)).
\label{11}
\end{equation}
Another example is given by taking\label{def8} $K_{1}$ and $K_{2}$ as the
projections to the corresponding subspaces $\mathcal{H}_{1},\mathcal{H}_{2}$
of the orthogonal sum $L^{2}(G)=\mathcal{H}_{1}\oplus \mathcal{H}_{2}$.

In \cite{FO} we used the first example in order to describe a teleportation
model where Bob performs his experiments on the same ensemble of the systems
like Alice. Further we used a special case of the second example in order to
describe a teleportation model where Bob and Alice are spatially separated
(cf. section 5 of \cite{FO}).

\section{Previous results on teleportation}

\label{sec2} Let us review some results obtained in \cite{FO}. We fix an ONS 
$\{g_{1},\ldots ,g_{N}\}\subseteq L^{2}(G)$, operators $K_{1},K_{2}$ on $%
L^{2}(G)$ with (\ref{5}), an unitary operator $T$ on $L^{2}(G)$, and $d>0$.
We assume 
\begin{equation}
TK_{1}g_{k}=K_{2}g_{k}\quad (k=1,\ldots ,N),  \label{12}
\end{equation}
\begin{equation}
\langle K_{1}g_{k},K_{1}g_{j}\rangle =0\quad (k\not{=}j;\;k,j=1\ldots ,N),
\label{13}
\end{equation}
Using (\ref{11}) and (\ref{12}) we get 
\begin{equation}
\Vert K_{1}g_{k}\Vert ^{2}=\Vert K_{2}g_{k}\Vert ^{2}=\frac{1}{2}.
\label{14}
\end{equation}
From (\ref{12}) and (\ref{13}) we get 
\begin{equation}
\langle K_{2}g_{k},\,K_{2}g_{j}\rangle =0\quad (k\neq j\,;\;k,j=1,\ldots ,N).
\label{15}
\end{equation}
The state of Alice asked to teleport is of the type 
\begin{equation}
\rho =\sum\limits_{s=1}^{N}\lambda _{s}|\Phi _{s}\rangle \langle \Phi _{s}|,
\label{16}
\end{equation}
where 
\begin{equation}
|\Phi _{s}\rangle =\sum\limits_{j=1}^{N}c_{sj}|\mathrm{exp}\;(aK_{1}g_{j})-%
\mathrm{exp}\;(0)\rangle \quad \left(
\sum\limits_{j}|c_{sj}|^{2}=1;s=1,\ldots ,N\right)  \label{17}
\end{equation}
and $a=\sqrt{d}$. One easily checks that $(|\mathrm{exp}\;(aK_{1}g_{j})-%
\mathrm{exp}\;(0)\rangle )_{j=1}^{N}$ and $(|\mathrm{exp}\;aK_{2}g_{j})-%
\mathrm{exp}\;(0)\rangle )_{j=1}^{N}$ are ONS in $\mathcal{M}$. The set $%
\left\{ \Phi _{s};s=1,\ldots ,N\right\} $ makes the $N$-dimensional Hilbert
space $\mathcal{H}_{1}$ defining an input state teleported by Alice. We may
include the vaccum state $|\mathrm{exp}\;(0)\rangle $ to define $\mathcal{H}%
_{1},$ however we take the $N$-dimensional Hilbert space $\mathcal{H}_{1}$
as above because of computational simplicity.

In order to achieve that $(|\Phi _{s}\rangle )_{s=1}^{N}$ is still an ONS in 
$\mathcal{M}$ we assume 
\begin{equation}
\sum\limits_{j=1}^{N}\bar{c}_{sj}c_{kj}=0\quad (j\neq k\,;\;j,k=1,\ldots
,N)\,.  \label{18}
\end{equation}
Denote $c_{s}=[c_{s1,\ldots ,}c_{sN}]\in \Bbb{C}^{N}$, then $%
(c_{s})_{s=1}^{N}$ is an CONS in $\Bbb{C}^{N}$.

Let $(b_{n})_{n=1}^{N}$ be a sequence in $\Bbb{C}^{N}$, 
\[
b_{n}=[b_{n1,\ldots ,}b_{nN}] 
\]
with properties 
\begin{equation}
|b_{nk}|=1\quad (n,k=1,\ldots ,N),  \label{19}
\end{equation}
\begin{equation}
\langle b_{n}\,,\;b_{j}\rangle =0\quad (n\neq j\,;\;n,j=1,\ldots ,N).
\label{20}
\end{equation}
Now, for each $m,n\left( =1,\ldots ,N\right) ,$ we have unitary operators $%
U_{m},B_{n}$ on $\mathcal{M}$ given by 
\begin{equation}
B_{n}|\mathrm{exp}\;(aK_{1}g_{j})-\mathrm{exp}\;(0)\rangle =b_{nj}|\mathrm{%
exp}\;(aK_{1}g_{j})-\mathrm{exp}\;(0)\rangle \quad (j=1,\ldots ,N)
\label{21}
\end{equation}
\begin{equation}
U_{m}|\mathrm{exp}\;(aK_{1}g_{j})-\mathrm{exp}\;(0)\rangle =|\mathrm{exp}
\;(aK_{1}g_{j\oplus m})-\mathrm{exp}\;(0)\rangle \quad (j=1,\ldots ,N)
\label{22}
\end{equation}

\subsection{A perfect teleportation}

Then Alice's measurements are performed with projection 
\begin{equation}
F_{nm}=|\xi _{nm}\rangle \langle \xi _{nm}|\quad (n,m=1,\ldots ,N)
\label{23}
\end{equation}
given by 
\begin{equation}
|\xi _{nm}\rangle =\frac{1}{\sqrt{N}}\sum\limits_{j=1}^{N}b_{nj}|\mathrm{exp}
\;(aK_{1}g_{j})-\mathrm{exp}\;(0)>\otimes |\;\mathrm{exp}\;(aK_{1}g_{j\oplus
m})-\mathrm{exp}\;(0)\rangle ,  \label{24}
\end{equation}
where $j\oplus m:=j+m(\mathrm{mod}\;N)$. \newline

One easily checks that $(|\xi _{nm}\rangle )_{n,m=1}^{N}$ is an ONS in $%
\mathcal{M}^{\otimes 2}$. Further, the state vector $|\xi \rangle $ of the
entangled state $\sigma =|\xi \rangle \langle \xi |$ is given by 
\begin{equation}
|\xi \rangle =\frac{1}{\sqrt{N}}\sum\limits_{k}|\mathrm{exp}\;(aK_{1}g_{k})-%
\mathrm{exp}\;(0)\rangle \otimes |\mathrm{exp}\;(aK_{2}g_{k})-\mathrm{exp}%
\;(0)\rangle \,.  \label{25}
\end{equation}
In \cite{FO} we proved the following theorem.

\begin{ftheo}
\label{def12} For each $n,m=1,\ldots ,N$, define a channel $\Lambda _{nm}$
by 
\begin{equation}
\Lambda _{nm}(\rho ):=\mathrm{tr}_{12}\frac{\left( F_{nm}\otimes 1\right)
(\rho \otimes \sigma )\left( F_{nm}\otimes \mathbf{1}\right) }{\mathrm{tr}%
_{123}\left( F_{nm}\otimes 1\right) \left( {\rho }\otimes \sigma \right)
\left( F_{nm}\otimes \mathbf{1}\right) }\quad (\rho \text{ normal state on }%
\mathcal{M})  \label{26}
\end{equation}
Then we have for all states $\rho $ on $M$ with (\ref{16}) and (\ref{17}) 
\begin{equation}
\Lambda _{nm}(\rho )=\left( \Gamma (T)U_{m}B_{n}^{*}\right) \rho \left(
\Gamma (T)U_{m}B_{n}^{*}\right) ^{*}  \label{27}
\end{equation}
\end{ftheo}

\begin{frem}
\label{def13} Using the operators $B_{n},U_{m},\Gamma (T),$ the projections $%
F_{nm}$ are given by unitary transformations of the entangled state $\sigma $
: 
\begin{eqnarray}
F_{nm} &=&\left( B_{n}\otimes U_{m}\Gamma (T^{*})\right) \sigma \left(
B_{n}\otimes U_{m}\Gamma (T^{*})\right) ^{*}  \label{28} \\
&&\text{or}\hspace*{2cm}  \nonumber \\
|\xi _{nm}\rangle &=&\left( B_{n}\otimes U_{m}\Gamma (T^{*})\right) |\xi
\rangle .  \nonumber
\end{eqnarray}
\end{frem}

If Alice performs a measurement according to the following selfadjoint
operator 
\[
F=\sum\limits_{n,m=1}^{N}z_{nm}F_{nm} 
\]
with $\{z_{nm}|n,m=1,\ldots ,N\}\subseteq \mathbf{R}-\{0\},$ then she will
obtain the value $z_{nm}$ with probability $1/N^{2}$. The sum over all this
probabilities is $1$, so that the teleportation model works perfectly.

Before stating some fundamental results in \cite{FO} for non-perfect case,
we note that our perfect teleportation is obviously treated in general
finite Hilbert spaces $\mathcal{H}_{k}\mathcal{\ }\left( k=1,2,3\right) $
same as usual one \cite{AO2}. Moreover, our teleportation scheme can be a
bit generalized by introducing the entagled state $\sigma _{12}$ on $%
\mathcal{H}_{1}\otimes \mathcal{H}_{2}$ defining the projections $\left\{
F_{nm}\right\} $ by the unitary operators $B_{n},U_{m}.$ We here discuss the
perfect teleportation on general Hilbert spaces $\mathcal{H}_{k}\mathcal{\ }%
\left( k=1,2,3\right) .$ Let $\left\{ \xi _{j}^{k};j=1,\cdots ,N\right\} $
be CONS of the Hilbert space $\mathcal{H}_{k}\mathcal{\ }\left(
k=1,2,3\right) .$ Define the entangled states $\sigma _{12}$ and $\sigma
_{23}$ on $\mathcal{H}_{1}\otimes \mathcal{H}_{2}$ and $\mathcal{H}
_{2}\otimes \mathcal{H}_{3},$ respectively, such as 
\[
\sigma _{12}=|\xi _{12}\rangle \left\langle \xi _{12}\right| ,\text{ }\sigma
_{23}=|\xi _{23}\rangle \left\langle \xi _{23}\right| 
\]
with $\xi _{12}\equiv \frac{1}{\sqrt{N}}\sum_{j=1}^{N}\xi _{j}^{1}\otimes
\xi _{j}^{2}$ and $\xi _{23}\equiv \frac{1}{\sqrt{N}}\sum_{j=1}^{N}\xi
_{j}^{2}\otimes \xi _{j}^{3}.$ By a sequence $\left\{ b_{n}=[b_{n1,\ldots
,}b_{nN}];n=1,\cdots ,N\right\} $ in $\Bbb{C}^{N}$ with the properties (19)
and (20), we define the unitary operator $B_{n}$ and $U_{m}$ such as

\[
B_{n}\xi _{j}^{1}\equiv b_{nj}\xi _{j}^{1}(n,j=1,\cdots ,N)\text{ and}%
U_{m}\xi _{j}^{2}\equiv \xi _{j\oplus m}^{2}(n,j=1,\cdots ,N) 
\]
with $j\oplus m\equiv j+m$ (mod $N).$ Then the set $\left\{
F_{nm};n,m=1,\cdots ,N\right\} $ of the projections of Alice is given by

\[
F_{nm}=\left( B_{n}\otimes U_{m}\right) \sigma _{12}\left( B_{n}\otimes
U_{m}\right) ^{*} 
\]
and the teleportation channels $\left\{ \Lambda _{nm}^{*};n,m=1,\cdots
,N\right\} $ are defined as

\[
\Lambda _{nm}(\rho ):=\mathrm{tr}_{12}\frac{\left( F_{nm}\otimes 1\right)
(\rho \otimes \sigma _{23})\left( F_{nm}\otimes \mathbf{1}\right) }{\mathrm{%
\ tr}_{123}\left( F_{nm}\otimes 1\right) \left( {\rho }\otimes \sigma
_{23}\right) \left( F_{nm}\otimes \mathbf{1}\right) }\quad (\rho \text{
normal state on }\mathcal{H}_{1}). 
\]
Finally the unitary keys $\left\{ W_{nm};n,m=1,\cdots ,N\right\} $ of Bob
are given as

\[
W_{nm}\xi _{j}^{1}=\overline{b}_{nj}\xi _{j\oplus m}^{3},\text{ }\left(
n,m=1,\cdots ,N\right) 
\]
by which we obtain the perfect teleportation

\[
\Lambda _{nm}(\rho )=W_{nm}\rho W_{nm}^{*}. 
\]
The above perfect teleportation is unique in the sense of unitary
equivalence.

\subsection{A non--perfect teleportation}

We will review a non-perfect teleportation model in which the probability
teleporting the state from Alice to Bob is less than $1$ and it depends on
the density parameter $d$ (may be the energy of the beams) of the coherent
vector.There, when $d=a^{2}$ tends to infinity, the probability tends to $1$%
. Thus the model can be considered as asymptotically perfect.

Take the normalized vector 
\begin{eqnarray}
|\eta \rangle := &&\frac{\gamma }{\sqrt{N}}\sum\limits_{k=1}^{N}|\mathrm{exp}
\;(ag_{k})\rangle  \label{29} \\
\text{with }\gamma := &&\left( \frac{1}{1+(N-1)e^{-d}}\right) ^{\frac{1}{2}
}=\left( \frac{1}{1+(N-1)e^{-a^{2}}}\right) ^{\frac{1}{2}}  \nonumber
\end{eqnarray}
and we replace in (\ref{26}) the entangled state $\sigma $ by 
\begin{eqnarray}
\tilde{\sigma}:= &&|\tilde{\xi}\rangle \langle \tilde{\xi}|  \label{30} \\
\tilde{\xi}:= &&\nu _{K_{1},K_{2}}(\eta )=\frac{\gamma }{\sqrt{N}}%
\sum\limits_{k=1}^{N}|\mathrm{exp}\;(aK_{1}g_{k})\rangle \otimes |\mathrm{exp%
}\;(aK_{2}g_{k})\rangle  \nonumber
\end{eqnarray}
Then for each $n,m=1,\ldots ,N,$ we get the channels on any normal state $%
\rho $ on $\mathcal{M}$ such as 
\begin{eqnarray}
\tilde{\Lambda}_{nm}(\rho ):= &&\mathrm{tr}_{12}\frac{\left( F_{nm}\otimes 
\mathbf{1}\right) \left( \rho \otimes \tilde{\sigma}\right) \left(
F_{nm}\otimes \mathbf{1}\right) }{\mathrm{tr}_{123}\left( F_{nm}\otimes 
\mathbf{1}\right) \left( \rho \otimes \tilde{\sigma}\right) \left(
F_{nm}\otimes \mathbf{1}\right) }\quad  \label{31} \\[0.12in]
\Theta _{nm}(\rho ):= &&\mathrm{tr}_{12}\frac{\left( F_{nm}\otimes
F_{+}\right) \left( \rho \otimes \tilde{\sigma}\right) \left( F_{nm}\otimes
F_{+}\right) }{\mathrm{tr}_{123}\left( F_{nm}\otimes F_{+}\right) \left(
\rho \otimes \tilde{\sigma}\right) \left( F_{nm}\otimes F_{+}\right) }\;,
\label{32}
\end{eqnarray}
where $F_{+}=\mathbf{1}-|\mathrm{exp}\;(0)\rangle \langle \mathrm{exp}
\;(0)|, $ i.e.., $F_{+}$ is the projection onto the space $\mathcal{M}_{+}$
of configurations having no vacuum part; 
\[
\mathcal{M}_{+}:=\{\psi \in \mathcal{M}|\;\Vert \mathrm{exp}\;(0)\rangle
\langle \mathrm{exp}\;(0)|\psi \Vert =0\} 
\]
One easily checks that 
\begin{equation}
\Theta _{nm}(\rho )=\frac{F_{+}\tilde{\Lambda}_{nm}(\rho )F_{+}}{\mathrm{tr}
\left( F_{+}\tilde{\Lambda}_{nm}(\rho )F_{+}\right) }  \label{33}
\end{equation}
that is, after receiving the state $\tilde{\Lambda}_{nm}(\rho )$ from Alice,
Bob has to omit the vacuum. \newline

From Theorem \ref{def12} it follows that for all $\rho $ with (\ref{16}) and
(\ref{17})

\[
\Lambda _{nm}(\rho )=\frac{F_{+}\Lambda _{nm}(\rho )F_{+}}{\mathrm{tr}
\;(F_{+}\Lambda _{nm}(\rho )F_{+})}. 
\]

This is not true if we replace $\Lambda _{nm}$ by $\tilde{\Lambda}_{nm}$,
namely, in general it does not hold 
\[
\Theta _{nm}(\rho )=\tilde{\Lambda}_{nm}(\rho ) 
\]

In \cite{FO} we proved the following theorem.

\begin{ftheo}
\label{def15} For all states $\rho $ on $\mathcal{M}$ with (\ref{16}) and (%
\ref{17}) and each pair $n,m\left( =1,\ldots ,N\right) ,$ we have 
\begin{equation}
\Theta _{nm}(\rho )=\left( \Gamma \left( T\right) U_{m}B_{n}^{*}\right) \rho
\left( \Gamma \left( T\right) U_{m}B_{n}^{*}\right) ^{*}\text{\quad or\quad }%
\Theta _{nm}(\rho )=\Lambda _{nm}(\rho )  \label{34}
\end{equation}
and 
\begin{equation}
\sum\limits_{n,m}p_{nm}(\rho )=\sum\limits_{n,m}\mathrm{tr}_{123}\left(
F_{nm}\otimes F_{+}\right) \left( \rho \otimes \tilde{\sigma}\right) \left(
F_{nm}\otimes F_{+}\right) =\frac{\left( 1-e^{-\frac{d}{2}}\right) ^{2}}{%
1+(N-1)e^{-d}}.  \label{35}
\end{equation}
\end{ftheo}

That is, the model works only asymptotically perfect in the sense of
condition (E2). With other words, in the case of high density (or energy) of
the considered beams the model works perfectly.

\section{Main results}

\label{sec3} The tools of the teleportation model considered in section~2.1
are the entangled state $\sigma $ and the family of projections $%
(F_{nm})_{n,m=1}^{N}$. In order to have a more handy model, in section~2.2.
we have replaced the entangled state $\sigma $ by another entangled state $%
\tilde{\sigma}$ given by the splitting of a superposition of certain
coherent states (\ref{30}). In addition now we are going to replace the
projectors $F_{nm}$ by projectors $\tilde{F}_{nm}$ defined as follows. 
\begin{equation}
\tilde{F}_{nm}:=\left( B_{n}\otimes U_{m}\Gamma (T)^{*}\right) \tilde{\sigma}%
\left( B_{n}\otimes U_{m}\Gamma (T)^{*}\right) ^{*}  \label{36}
\end{equation}
In order to make this definition precise we assume, in addition to (\ref{22}
), that is holds: 
\[
U_{m}\mathrm{exp}(0)=\mathrm{exp}(0)\quad (m=1,\ldots ,N) 
\]
Together with (\ref{22}) that implies 
\begin{equation}
U_{m}|\mathrm{exp}(aK_{1}g_{j})\rangle =|\mathrm{exp}(aK_{1}g_{j\oplus
m})\rangle \quad (m,j=1,\ldots ,N)  \label{37}
\end{equation}
Formally we have the same relation between $\tilde{\sigma}$ and $\tilde{F}%
_{nm}$ like the relation between $\sigma $ and $F_{nm}$ (cf. Remark \ref
{def13}). Further for each pair $n,m=1,\ldots ,N$ we define channels on
normal states on $\mathcal{M}$ such as 
\begin{equation}
\tilde{\Theta}_{nm}(\rho ):=\mathrm{tr}_{12}\frac{\left( \tilde{F}%
_{nm}\otimes F_{+}\right) \left( \rho \otimes \tilde{\sigma}\right) \left( 
\tilde{F}_{nm}\otimes F_{+}\right) }{\tilde{p}_{nm}(\rho )}  \label{38}
\end{equation}
where 
\begin{equation}
\tilde{p}_{nm}(\rho ):=\mathrm{tr}_{123}\left( \tilde{F}_{nm}\otimes
F_{+}\right) \left( \rho \otimes \tilde{\sigma}\right) \left( \tilde{F}%
_{nm}\otimes F_{+}\right)  \label{39}
\end{equation}
(cf. (\ref{33}), and (\ref{34})).\newline

In section 4, we will prove the following theorem.

\begin{ftheo}
\label{def17} For each state $\rho $ on $\mathcal{M}$ with (\ref{16}), and (%
\ref{17}), each pair $n,m(=1,\ldots ,N)$ and each bounded operator $A$ on $%
\mathcal{M}$ it holds 
\begin{equation}
|\mathrm{tr}\left( \tilde{\Theta}_{nm}(\rho )A\right) -\mathrm{tr}\left(
\Lambda _{nm}(\rho )A\right) |\le \frac{2e^{-\frac{d}{2}}}{\left( 1-e^{-%
\frac{d}{2}}\right) }\left( N^{2}+N\sqrt{N}+N\right)  \label{40}
\end{equation}
\begin{equation}
\left| \tilde{p}_{nm}(\rho )-\frac{1}{N^{2}}\right| \le e^{-\frac{d}{2}%
}\left( \frac{14}{N^{2}}+2+\frac{2}{\sqrt{N}}\right)  \label{41}
\end{equation}
\end{ftheo}

From Theorem \ref{def12} and $e^{-\frac{d}{2}}{\longrightarrow }0\left( d\to
+\infty \right) ,$ the theorem \ref{def17} means that our modified
teleportation model works asymptotically perfect (case of high density or
energy) in the sense of conditions (E1), and (E2).\newline

In order to obtain a deeper understanding of the whole procedure we are
going to discuss another representation of the projectors $\tilde{F}_{nm}$
and of the channels $\tilde{\Theta}_{nm}$. Starting point is again the
normalized vector $|\eta \rangle $ given by (\ref{29}). From (\ref{14}) we
obtain 
\begin{equation}
\Vert O_{\sqrt{2}}K_{1}g_{k}\Vert ^{2}=\Vert g_{k}\Vert ^{2},  \label{42}
\end{equation}
where $O_{f}$ denotes the operator of multiplication corresponding to the
number (or function) $f$ 
\begin{equation}
O_{f}\psi :=f\psi \quad \left( \psi \in L_{2}(G)\right)  \label{43}
\end{equation}
Furthermore (\ref{13}) implies 
\begin{equation}
\langle O_{f}K_{1}g_{k}\;,\;O_{f}K_{1}g_{j}\rangle =0\quad (k\neq j)
\label{44}
\end{equation}
From (\ref{42}), and (\ref{44}) follows that we have a normalized vector $|%
\tilde{\eta}\rangle $ given by 
\begin{equation}
|\tilde{\eta}\rangle :=\Gamma \left( O_{\sqrt{2}}K_{1}\right) |\eta \rangle =%
\frac{\gamma }{\sqrt{N}}\sum\limits_{k=1}^{N}|\mathrm{exp}\left( a\sqrt{2}%
K_{1}g_{k}\right) \rangle  \label{45}
\end{equation}

\begin{frem}
\label{def18} In the case of Example \ref{def7} we have 
\[
|\tilde{\eta}\rangle =|\eta \rangle 
\]
\end{frem}

Now let $V$ be the unitary operator on $\mathcal{M}\otimes \mathcal{M}$
characterized by

\begin{eqnarray}
&&V\left( \mathrm{exp}(f_{1})\otimes \mathrm{exp}(f_{2})\right)  \nonumber \\
&=&\mathrm{exp}\left( \frac{1}{\sqrt{2}}\left( f_{1}-f_{2}\right) \right)
\otimes \mathrm{\ exp}\left( \frac{1}{\sqrt{2}}\left( f_{1}+f_{2}\right)
\right) \left( f_{1},f_{2}\in L_{2}(G)\right)  \label{46}
\end{eqnarray}
On easily checks

\begin{eqnarray}
&&V^{*}\left( \mathrm{exp}(f_{1})\otimes \mathrm{exp}(f_{2})\right) 
\nonumber \\
&=&\mathrm{exp}\left( \frac{1}{\sqrt{2}}\left( f_{1}+f_{2}\right) \right)
\otimes \mathrm{exp}\left( \frac{1}{\sqrt{2}}\left( f_{2}-f_{1}\right)
\right) \left( f_{1},f_{2}\in L_{2}(G)\right)  \label{47}
\end{eqnarray}

\begin{frem}
\label{def19} $V$ describes a certain exchange procedure of particles (or
energy) between two systems or beams (cf. \cite{FFL3})
\end{frem}

Now, using (\ref{12}), (\ref{30}), (\ref{45}), and (47), resp. (46) one gets 
\begin{eqnarray}
\tilde{\xi} &=&\nu _{K_{1},K_{2}}(\eta )=(\mathbf{1}\otimes \Gamma
(T))V^{*}\left( |\mathrm{exp}(0)\rangle \otimes |\tilde{\eta}\rangle \right)
\label{48} \\
\tilde{\xi} &=&(\mathbf{1}\otimes \Gamma (T))V\left( |\tilde{\eta}\rangle
\otimes |\mathrm{exp}(0)\rangle \right)  \label{49}
\end{eqnarray}
and it follows 
\begin{eqnarray}
\tilde{\sigma} &=&|\tilde{\xi}\rangle \langle \tilde{\xi}|  \nonumber \\
&=&(\mathbf{1}\otimes \Gamma (T))V^{*}\left( |\mathrm{exp}(0)\rangle \langle 
\mathrm{exp}(0)|\otimes |\tilde{\eta}\rangle \langle \tilde{\eta}|\right)
\left( (\mathbf{1}\otimes \Gamma (T))V^{*}\right) ^{*}  \label{50}
\end{eqnarray}
\begin{equation}
\tilde{\sigma}=(\mathbf{1}\otimes \Gamma (T))V\left( |\tilde{\eta}\rangle
\langle \tilde{\eta}|\otimes |\mathrm{exp}(0)\rangle \langle \mathrm{\ exp}
(0)|\right) ((\mathbf{1}\otimes \Gamma (T))V)^{*}  \label{51}
\end{equation}
From the definition of $\tilde{F}_{nm}$ (\ref{36}) and (\ref{50}) it follows 
\begin{equation}
\tilde{F}_{nm}=\left( B_{n}\otimes U_{m}\right) V^{*}\left( |\mathrm{exp}
(0)\rangle \langle \mathrm{exp}(0)|\otimes |\tilde{\eta}\rangle \langle 
\tilde{\eta}|\right) \left( \left( B_{n}\otimes U_{m}\right) V^{*}\right)
^{*}  \label{52}
\end{equation}
Using (\ref{51}), and (\ref{52}) we obtain 
\begin{eqnarray}
&&\left( \tilde{F}_{nm}\otimes F_{+}\right) \left( \rho \otimes \tilde{\sigma%
}\right) \left( \tilde{F}_{nm}\otimes F_{+}\right) \quad (n,m=1,\ldots ,N) 
\nonumber \\
&=&\left( X_{nm}\otimes \mathbf{1}\right) W_{nm}\left( \rho \otimes |\tilde{%
\eta}\rangle \langle \tilde{\eta}|\otimes |\mathrm{exp}(0)\rangle \langle 
\mathrm{exp}(0)|\right) W_{nm}^{*}\left( X_{nm}\otimes \mathbf{1}\right) ^{*}
\label{53}
\end{eqnarray}
where 
\begin{equation}
\begin{split}
X_{nm}& :=\left( B_{n}\otimes U_{m}\right) V^{*}\qquad (n,m=1,\ldots ,N) \\
W_{nm}& :=\left( |\mathrm{exp}(0)\rangle \langle \mathrm{exp}(0)|\otimes |%
\tilde{\eta}\rangle \langle \tilde{\eta}|\otimes F_{+}\right) \left(
V\otimes \mathbf{1}\right) \left( B_{n}^{*}\otimes U_{m}^{*}\otimes \Gamma
(T)\right) \left( \mathbf{1}\otimes V\right)
\end{split}
\label{54}
\end{equation}
$X_{nm}$ and consequently $X_{nm}\otimes \mathbf{1}$ are unitary operators.
For that reason we get from (\ref{53})

\begin{eqnarray}
&&\mathrm{tr}_{123}\left( \tilde{F}_{nm}\otimes F_{+}\right) \left( \rho
\otimes \tilde{\sigma}\right) \left( \tilde{F}_{nm}\otimes F_{+}\right)
(n,m=1,\ldots ,N)  \nonumber \\
&=&\mathrm{tr}_{123}W_{nm}\left( \rho \otimes |\tilde{\eta}\rangle \langle 
\tilde{\eta}|\mathrm{exp}(0)\rangle \langle \mathrm{exp}(0)|\right)
W_{nm}^{*}  \label{55}
\end{eqnarray}
and 
\begin{eqnarray}
&&\mathrm{tr}_{12}\left( \tilde{F}_{nm}\otimes F_{+}\right) \left( \rho
\otimes \tilde{\sigma}\right) \left( \tilde{F}_{nm}\otimes F_{+}\right)  
\nonumber \\
&=&\mathrm{tr}_{12}W_{nm}\left( \rho \otimes |\tilde{\eta}\rangle \langle 
\tilde{\eta}|\otimes |\mathrm{exp}(0)\rangle \langle \mathrm{exp}(0)|\right)
W_{nm}^{*}  \label{56}
\end{eqnarray}
Now from (\ref{38}), (\ref{39}), (\ref{55}) and (\ref{56}) it follows 
\begin{eqnarray}
\tilde{p}_{nm}(\rho ) &=&\mathrm{tr}_{123}W_{nm}\left( \rho \otimes |\tilde{%
\eta}\rangle \langle \tilde{\eta}|\otimes |exp(0)\rangle \langle \mathrm{exp}%
(0)|\right) W_{nm}^{*}  \label{57} \\
\tilde{\Theta}_{nm}(\rho ) &=&tr_{12}\frac{W_{nm}\left( \rho \otimes |\tilde{%
\eta}\rangle \langle \tilde{\eta}|\otimes |\mathrm{exp}(0)\rangle \langle 
\mathrm{exp}(0)|\right) W_{nm}^{*}}{\mathrm{tr}_{123}W_{nm}\left( \rho
\otimes |\tilde{\eta}\rangle \langle \tilde{\eta}|\otimes |\mathrm{exp}%
(0)\rangle \langle \mathrm{exp}(0)|\right) W_{nm}^{*}}  \label{58}
\end{eqnarray}
According to (\ref{57},\ref{58}) and (\ref{54}), the procedure of the
special teleportation model can be expressed in the following steps.\newpage 

\begin{tabular}{lc}
\textbf{Step 0 --}\textbf{\textit{initial state}} & $s_{\mathrm{in}}(\rho
)=\rho \otimes |\tilde{\eta}\rangle \langle \tilde{\eta}|\otimes |\mathrm{exp%
}(0)\rangle \langle \mathrm{exp}(0)|$ \\ 
$\rho $--the unknown state & $|$ \\ 
Alice want to teleport & $|$ \\ 
$|\mathrm{exp}(0)\rangle \langle \mathrm{exp}(0)|$--vacuum state, & $|$ \\ 
Bobs state at the beginning & $|$ \\ 
& $|$ \\ 
\textbf{Step 1 --}\textbf{\textit{Transformation according to}} & $\mathbf{1}
\otimes V$ \\ 
that means: splitting of & $|$ \\ 
the state $|\tilde{\eta}\rangle \langle \tilde{\eta}|$ & $|$ \\ 
& $|$ \\ 
\textbf{Step 2 --}\textbf{\textit{Transformation according to}} & $%
B_{n}^{*}\otimes U_{m}^{*}\otimes \Gamma (T)$ \\ 
& $|$ \\ 
& $|$ \\ 
\textbf{Step 3 --}\textbf{\textit{Transformation according to}} & $V\otimes 
\mathbf{1}$ \\ 
exchange of particles (or energy) & $|$ \\ 
between the first and the second & $|$ \\ 
part of the system & $|$ \\ 
& $|$ \\ 
\textbf{Step 4 --}\textbf{\textit{measurement according to}} & $|\mathrm{exp}
(0)\rangle \langle \mathrm{exp}(0)|\otimes |\tilde{\eta}\rangle \langle 
\tilde{\eta}|\otimes F_{+}$ \\ 
checking for & $|$ \\ 
- first part in the vacuum? & $|$ \\ 
- in the third part is no vacuum? & $|$ \\ 
- second part reconstructed? & $|$ \\ 
& $\downarrow $ \\ 
&  \\ 
\hspace*{\fill} \textbf{\textit{Final state}} $s_{\mathrm{fin}}(\rho )$ & $=%
\frac{W_{nm}\left( s_{\mathrm{in}}(\rho )\right) W_{nm}^{*}}{\mathrm{tr}%
_{123}W_{nm}\left( s_{\mathrm{in}}(\rho )\right) W_{nm}^{*}}$%
\end{tabular}

\vspace*{5mm} Now from (\ref{57}) we get $\tilde{\Theta}_{nm}(\rho )=\mathrm{%
\ tr}_{12}\;s_{\mathrm{fin}}(\rho )$. Thus theorem \ref{def17} means that in
the case of high density (or energy) $d$ we have approximately ($\rho $ with
(\ref{16}), and (\ref{17})) 
\[
\mathrm{tr}_{12}\;s_{\mathrm{fin}}(\rho )=\left( \Gamma
(T)U_{m}B_{n}^{*}\right) \rho \left( \Gamma (T)U_{m}B_{n}^{*}\right) ^{*} 
\]
The proof of theorem \ref{def17} shows that we have even more, namely it
holds (approximately) 
\begin{equation}
s_{\mathrm{fin}}(\rho )=|\mathrm{exp}(0)\rangle \langle \mathrm{exp}
(0)|\otimes |\tilde{\eta}\rangle \langle \tilde{\eta}|\otimes \left( \Gamma
(T)U_{m}B_{n}^{*}\right) \rho \left( \Gamma (T)U_{m}B_{n}^{*}\right) ^{*}
\label{59}
\end{equation}
Adding in our scheme the following step\newline

\noindent \textbf{Step 5 --}\textbf{\textit{Transformation}} \hspace*{40mm}$%
\mathbf{1}\otimes \mathbf{1}\otimes \left( \Gamma (T)U_{m}B_{n}^{*}\right)
^{*}$\newline
(that means Bob uses the key provided to him)\newline

Then $s_{\mathrm{fin}}(\rho )$ will change into the new final state 
\[
|\mathrm{exp}(0)\rangle \langle \mathrm{exp}(0)|\otimes |\tilde{\eta}\rangle
\langle \tilde{\eta}|\otimes \rho 
\]
Summarizing one can describe the effect of the procedure (for large $d$!) as
follows: At the beginning Alice has (e.~g., can control) a state $\rho $,
and Bob has the vacuum state (e.~g., can control nothing). After the
procedure Bob has the state $\rho $ and Alice has the vacuum. Furthermore
the teleportation mechanism is ready for the next teleportation (e.~g. $|%
\tilde{\eta}\rangle \langle \tilde{\eta}|$ is reproduced in the course of
teleportation).\newline

We have considered three different models (cf. sections 2.1, 2.2, 2.3). Each
of them is a special case of a more general concept we are going to describe
in the following:

Let $H_{1},H_{2}$ be $N$--dimensional subspaces of $\mathcal{M}_{+}$ such
that $\Gamma (T)$ maps $H_{1}$ onto $H_{2}$, and $H_{1}$ is invariant with
respect to the unitary transformations $B_{n}$, $U_{m}$ $(n,m=1,\ldots ,N)$.%
\newline

Further let $\sigma _{1}$, $\sigma _{2}$ be projections of the type 
\[
\sigma _{k}=|\xi _{k}\rangle \langle \xi _{k}|\quad ,\quad \xi _{k}\in
\left( H_{1}\oplus \mathcal{M}_{0}\right) \otimes \left( H_{2}\otimes 
\mathcal{M}_{0}\right) \quad (k=1,2) 
\]
where $\mathcal{M}_{0}$ is the orthogonal complement of $\mathcal{M}_{+}$,
e.~g., $\mathcal{M}_{0}$ is the one-dimensional subspace of $\mathcal{M}$
spanned by the vacuum vector $|\mathrm{exp}(0)\rangle $.\newline

Now for each $n,m=1,\ldots ,N$ and each pair $\sigma _{1}$, $\sigma _{2}$ we
define a channel $\Omega _{nm}^{\sigma _{1},\sigma _{2}}$ from the set of
all normal states $\rho $ on $H_{1}$ into the set of all normal states on $%
\mathcal{M}_{+}$ 
\[
\Omega _{nm}^{\sigma _{1}\sigma _{2}}(\rho ):=\mathrm{tr}_{12}\frac{\left(
F_{nm}^{\sigma _{1}}\otimes F_{+}\right) \left( \rho \otimes \sigma
_{2}\right) \left( F_{nm}^{\sigma _{1}}\otimes F_{+}\right) }{\mathrm{tr}%
_{123}\left( F_{nm}^{\sigma _{1}}\otimes F_{+}\right) \left( \rho \otimes
\sigma _{2}\right) \left( F_{nm}^{\sigma _{1}}\otimes F_{+}\right) } 
\]
where 
\[
F_{nm}^{\sigma _{1}}:=\left( B_{n}\otimes U_{m}\Gamma (T^{*})\right) \sigma
_{1}\left( B_{n}\otimes U_{m}\Gamma (T^{*})\right) ^{*} 
\]
In this paper we have considered the situation where $H_{1}$ is spanned by
the ONS 
\[
\left( |\mathrm{exp}(aK_{1}g_{k})-\mathrm{exp}(0)\rangle \right) _{k=1}^{N} 
\]
and $H_{2}$ is spanned by the ONS 
\[
\left( |\mathrm{exp}(aK_{2}g_{k})-\mathrm{exp}(0)\rangle \right) _{k=1}^{N} 
\]
Further the model discussed in section 2.2 corresponds to the special case $%
\sigma _{1}=\sigma _{2}=\sigma $, e.~g. 
\[
\Lambda _{nm}=\Omega _{nm}^{\sigma \sigma }\qquad (n,m=1,\ldots ,N) 
\]
(perfect in the sense of conditions (E1) and (E2)).\newline

The model discussed in section 2.2 corresponds to the special case $\sigma
_{1}=\sigma \neq \sigma _{2}=\tilde{\sigma}$, e.~g. 
\[
\Theta _{nm}=\Omega _{nm}^{\sigma \tilde{\sigma}} 
\]
(perfect in the sense of (E1), and only asymptotically perfect in the sense
of (E2)).\newline

Finally the model from this section corresponds to the special case $\sigma
_{1}=\sigma _{2}=\tilde{\sigma}$, e.~g. 
\[
\tilde{\Theta}_{nm}=\Omega _{nm}^{\tilde{\sigma}\tilde{\sigma}} 
\]
(non-perfect, neither (E2) nor (E1) hold, but asymptotically perfect in the
sense of both conditions)

\section{Proof of Theorem \ref{def17}}

\label{sec4} From (\ref{14}) we get 
\begin{equation}
\Vert \mathrm{exp}\left( aK_{s}g_{j}\right) -\mathrm{exp}(0)\Vert ^{2}=e^{%
\frac{a^{2}}{2}}-1\qquad (s=1,2;j=1,\ldots ,N)  \label{60}
\end{equation}
\begin{equation}
\Vert \mathrm{exp}\left( aK_{s}g_{j}\right) \Vert =e^{\frac{a^{2}}{2}}\qquad
(s=1,2;j=1,\ldots ,N)  \label{61}
\end{equation}
Using (\ref{46}), (\ref{60}) and (\ref{61}) one easily checks 
\begin{eqnarray}
&&V\left( |\mathrm{exp}\left( aK_{1}g_{j}\right) \right) -\mathrm{exp}
(0)\rangle \otimes |\mathrm{exp}\left( aK_{1}g_{k}\right) \rangle  \label{62}
\\
&&=\left( \left( e^{\frac{a^{2}}{2}}-1\right) e^{\frac{a^{2}}{2}}\right) ^{-%
\frac{1}{2}}\left[ \mathrm{exp}\left( \frac{a}{\sqrt{2}}K_{1}\left(
g_{j}-g_{k}\right) \right) \otimes \mathrm{exp}\left( \frac{a}{\sqrt{2}}%
K_{1}\left( g_{j}+g_{k}\right) \right) \right.  \nonumber \\
&&-\left. \mathrm{exp}\left( -\frac{a}{\sqrt{2}}K_{1}\left( g_{k}\right)
\right) \otimes \mathrm{exp}\left( \frac{a}{\sqrt{2}}K_{1}g_{k}\right)
\right]  \nonumber \\
&&(k,j=1,\ldots ,N)  \nonumber
\end{eqnarray}

\begin{flemma}
\label{def20} Put for $j,k=1,\ldots ,N$ 
\[
\alpha _{jk}:=\langle |\mathrm{exp}(0)\rangle \otimes |\tilde{\eta}\rangle
\;,\;V\left( |\mathrm{exp}\left( aK_{1}g_{j}\right) -\mathrm{exp}(0)\rangle
\otimes |\mathrm{exp}\left( aK_{1}g_{k}\right) \rangle \right) \rangle 
\]
Then it holds for $j,k=1,\ldots ,N$ 
\begin{equation}
\alpha _{jk}=\left( \left( 1-e^{-\frac{a^{2}}{2}}\right) e^{-a^{2}}\right) ^{%
\frac{1}{2}}\frac{\gamma }{\sqrt{N}}\qquad (k\neq j)  \label{63}
\end{equation}
\begin{equation}
\alpha _{jj}=\left( 1-e^{-\frac{a^{2}}{2}}\right) ^{\frac{1}{2}}\frac{\gamma 
}{\sqrt{N}}  \label{64}
\end{equation}
\end{flemma}

\noindent \textbf{Proof: }We have 
\begin{equation}
\langle \mathrm{exp}(0)\;,\;\mathrm{exp}(f)\rangle =1\qquad \left( f\in
L_{2}(G)\right)  \label{65}
\end{equation}
Using (\ref{62}), (\ref{65}), and (\ref{45}) we get for $j,k=1,\ldots ,N$ 
\begin{eqnarray}
&&\alpha _{jk}=\left( \left( e^{\frac{a^{2}}{2}}-1\right) e^{\frac{a^{2}}{2}
}\right) ^{-\frac{1}{2}}\frac{\gamma }{\sqrt{N}}\sum\limits_{s=1}^{N}\left[
\langle |\mathrm{exp}(\sqrt{2}\,aK_{1}g_{s})\rangle \;,\mathrm{exp}\left( 
\frac{a}{\sqrt{2}}K_{1}\left( g_{j}+g_{k}\right) \right) \rangle \right. 
\nonumber \\
&&\left. -\langle |\mathrm{exp}\left( \sqrt{2}\,aK_{1}g_{s}\right) \rangle
\;,\;\mathrm{exp}\left( \frac{a}{\sqrt{2}}K_{1}\left( g_{k}\right) \right)
\rangle \right]  \label{66}
\end{eqnarray}
We have 
\begin{equation}
\langle \mathrm{exp}(f_{1})\;,\;\mathrm{exp}(f_{2})\rangle =e^{\langle
f_{1},f_{2}\rangle }\qquad \left( f_{1},f_{2}\in L_{2}(G)\right)  \label{67}
\end{equation}
Using (\ref{13}) and (\ref{67}) we obtain 
\begin{eqnarray}
&&0=\langle \mathrm{exp}\left( \sqrt{2}\,aK_{1}g_{s}\right) \,,\,\mathrm{exp}%
\left( \frac{a}{\sqrt{2}}K_{1}\left( g_{j}+g_{k}\right) \right) \rangle
\qquad (s\neq j)  \nonumber \\
&&-\langle \mathrm{exp}\left( \sqrt{2}\,aK_{1},g_{s}\,,\,\mathrm{exp}\left( 
\frac{a}{\sqrt{2}}K_{1}g_{k}\right) \right) \rangle  \label{68}
\end{eqnarray}
From (\ref{61}), (\ref{66}), (\ref{67}), and (\ref{68}) it follows 
\begin{equation}
\alpha _{jk}=\left( \left( e^{\frac{a^{2}}{2}}-1\right) e^{\frac{a^{2}}{2}%
}e^{a^{2}}\right) ^{-\frac{1}{2}}\frac{\gamma }{\sqrt{n}}\left(
e^{a^{2}\langle K_{1}g_{j}\,,\,K_{1}(g_{j}+g_{k})\rangle }-e^{a^{2}\langle
K_{1}g_{j}\,,\,K_{1}g_{k}\rangle }\right)  \label{69}
\end{equation}
Now (\ref{13}) and (\ref{14}) implies 
\begin{equation}
\langle K_{1}g_{j}\,,\,K_{1}g_{k}\rangle =\frac{1}{2}\delta _{jk}  \label{70}
\end{equation}
For that reason (\ref{63}), and (\ref{64}) follow from (\ref{69}). $%
\blacksquare $\newline

In the following we fix a pair $n,m\in\{1,\ldots,N\}$.

\begin{frem}
\label{def21} Without loss of generality we can assume 
\begin{equation}
B_{n}=\mathbf{1}  \label{71}
\end{equation}
That we can explain as follows:\newline
Using (\ref{57},\ref{58}), (\ref{59}), and (\ref{54}) we obtain in the case (%
\ref{71}) 
\begin{eqnarray*}
\tilde{\Theta}_{km}(\rho ) &=&\tilde{\Theta}_{nm}\left( B_{k}^{*}\rho
B_{k}\right) \qquad (k=1,\ldots ,N) \\
\tilde{p}_{km}(\rho ) &=&\tilde{p}_{nm}\left( B_{k}^{*}\rho B_{k}\right)
\qquad (k=1,\ldots ,N)
\end{eqnarray*}
On the other hand from theorem \ref{def12} follows that in the case (\ref{71}
) for all states $\rho $ with (\ref{16}) and (\ref{17}) it holds 
\[
\Lambda _{km}(\rho )=\Lambda _{nm}\left( B_{k}^{*}\rho B_{k}\right) \qquad
(k=1,\ldots ,N) 
\]
Finally it is easy to show that $B_{k}^{*}\rho B_{k}$ fulfills (\ref{16}),
and (\ref{17}) if the state $\rho $ fulfills (\ref{16}) and (\ref{17}).%
\newline
For that reasons theorem \ref{def17} would be proved if we could prove (\ref
{40}), and (\ref{41}) on the assumption that we have (\ref{71}).\newline
\end{frem}

Now from (\ref{30}), (\ref{49}), and (\ref{37}) we get 
\begin{equation}  \label{72}
\left(U_m^*\otimes\Gamma(T)\right)V\left(|\tilde{\eta}\rangle\otimes|\mathrm{%
\ exp}(0)\rangle\right) =\frac{\gamma}{\sqrt{N}}\sum\limits_{k=1}^N|\mathrm{%
\ \ exp}\left(aK_1g_k\right)\rangle\otimes|\mathrm{exp} (aK_2g_{k\oplus
m})\rangle
\end{equation}

\begin{flemma}
\label{def22} Put for $s=1,\ldots ,N$ 
\begin{eqnarray}
&&\beta _{s}:=\left( \left( |\mathrm{exp}(0)\rangle \langle \mathrm{exp}
(0)|\otimes |\tilde{\eta}\rangle \langle \tilde{\eta}|\right) V\otimes 
\mathbf{1}\right) \left( \mathbf{1}\otimes U_{m}^{*}\otimes \Gamma
(T)\right) (\mathbf{1}\otimes V)|\Psi _{s}\rangle \otimes |\tilde{\eta}
\rangle \otimes  \nonumber \\
&&|\mathrm{exp}(0)\rangle \langle \mathrm{exp}(0)|\qquad (s=1,\ldots ,N)
\label{73}
\end{eqnarray}
Then it holds 
\begin{eqnarray}
&&\beta _{s}=\frac{\gamma ^{2}}{N}\left( 1-e^{-\frac{a^{2}}{2}}\right) ^{%
\frac{1}{2}}|\mathrm{exp}(0)\rangle \otimes |\tilde{\eta}\rangle \otimes
\left( \left( 1-e^{-\frac{a^{2}}{2}}\right) \sum\limits_{j=1}^{N}c_{sj}|%
\mathrm{exp}\left( aK_{2}g_{j\oplus m}\right) \rangle \right.  \nonumber \\
&&\left. +e^{-\frac{a^{2}}{2}}\sum\limits_{j=1}^{N}c_{sj}\sum%
\limits_{k=1}^{N}|\mathrm{exp}\left( aK_{2}g_{k}\right) \rangle \right)
\label{74}
\end{eqnarray}
\end{flemma}

\noindent \textbf{Proof: }From (\ref{17}), (\ref{72}), and (\ref{73}) we get 
\begin{eqnarray}
&&\beta _{s}=\sum\limits_{j=1}^{N}c_{sj}\frac{\gamma }{\sqrt{N}}
\sum\limits_{k=1}^{N}\left[ \left( |\mathrm{exp}(0)\rangle \langle \mathrm{\
exp}(0)|\otimes |\tilde{\eta}\rangle \langle \tilde{\eta}|\right) V\left( |%
\mathrm{exp}\left( aK_{1}g_{j}\right) -\mathrm{exp}(0)\rangle \right. \right.
\nonumber \\
&&\left. \left. \otimes |\mathrm{exp}\left( aK_{1}g_{k}\right) \rangle
\right) \right] \otimes |\mathrm{exp}\left( aK_{2}g_{k\oplus m}\right)
\rangle  \label{75}
\end{eqnarray}
Further we have 
\begin{eqnarray}
& &\left( |\mathrm{exp}(0)\rangle \langle \mathrm{exp}(0)|\otimes |\tilde{%
\eta}\rangle \langle \tilde{\eta}|\right) V\left( |\mathrm{exp}\left(
aK_{1}g_{j}\right) \right) -\mathrm{exp}(0)\rangle \otimes |\mathrm{exp}
\left( aK_{1}g_{k}\right) \rangle  \nonumber \\
& &=|\mathrm{exp}(0)\rangle \otimes |\tilde{\eta}\rangle \left\langle |%
\mathrm{\ exp}(0)\rangle \otimes |\tilde{\eta}\rangle \;,\;V\left( |\mathrm{%
exp}\left( aK_{1}g_{j}\right) -\mathrm{exp}(0)\rangle \otimes |\mathrm{exp}
\left( aK_{1}g_{k}\right) \rangle \right) \right\rangle  \nonumber \\
&& \hfill (j,k=1,\ldots ,N)  \label{76}
\end{eqnarray}
Using Lemma 4.1, (\ref{75}), and (\ref{76}) we obtain 
\begin{eqnarray*}
&&\beta _{s}=\frac{\gamma ^{2}}{N}\left( 1-e^{-\frac{a^{2}}{2}}\right) ^{%
\frac{1}{2}}|\mathrm{exp}(0)\rangle \otimes |\tilde{\eta}\rangle \otimes
\sum\limits_{j=1}^{N}c_{sj}|\mathrm{exp}\left( aK_{2}g_{j\oplus m}\right)
\rangle \\
&&+\frac{\gamma ^{2}}{N}\left( \left( 1-e^{-\frac{a^{2}}{2}}\right)
e^{-a^{2}}\right) ^{\frac{1}{2}}|\mathrm{exp}(0)\rangle \otimes |\tilde{\eta}
\rangle \otimes \left( \sum\limits_{j}\sum\limits_{k\neq j}c_{sj}|\mathrm{exp%
}\left( aK_{2}g_{k\oplus m}\right) \right) \rangle
\end{eqnarray*}
That implies (\ref{74}). $\blacksquare $\newline

Now we put 
\begin{equation}
|\Psi _{0}\rangle :=\frac{1}{\sqrt{N}}\sum\limits_{j=1}^{N}|\mathrm{exp}%
\left( aK_{1}g_{j}\right) -\mathrm{exp}(0)\rangle  \label{77}
\end{equation}
Since 
\[
F_{+}=\mathbf{1}-|\mathrm{exp}(0)\rangle \langle \mathrm{exp}(0)|, 
\]
one easily checks 
\begin{eqnarray}
&&F_{+}|\mathrm{exp}\left( aK_{r}g_{k}\right) \rangle =\left( 1-e^{-\frac{%
a^{2}}{2}}\right) ^{\frac{1}{2}}|\mathrm{exp}\left( aK_{r}g_{k}\right) -%
\mathrm{\exp }(0)\rangle  \nonumber \\
&&(r=1,2;\;k=1,\ldots ,m)  \label{78}
\end{eqnarray}
Using (\ref{77}), and (\ref{78}) we obtain 
\begin{eqnarray}
F_{+}\left( \sum\limits_{k=1}^{N}|\mathrm{exp}\left( aK_{2}g_{k}\right)
\rangle \right) &=&\left( 1-e^{-\frac{a^{2}}{2}}\right) ^{\frac{1}{2}}\sqrt{N%
}\,U_{m}\Gamma (T)|\Psi _{0}\rangle  \label{79} \\
&=&\left( 1-e^{-\frac{a^{2}}{2}}\right) ^{\frac{1}{2}}\sqrt{N}\,\Gamma
(T)U_{m}|\Psi _{0}\rangle  \nonumber
\end{eqnarray}
Using the same arguments we get 
\begin{eqnarray}
F_{+}\left( \sum\limits_{j=1}^{N}c_{sj}|\mathrm{exp}\left( aK_{2}g_{j\oplus
m}\right) \rangle \right) &=&\left( 1-e^{-\frac{a^{2}}{2}}\right) ^{\frac{1}{%
2}}U_{m}\Gamma (T)|\Psi _{s}\rangle \quad (s=1,\ldots ,N)  \label{80} \\
&=&\left( 1-e^{-\frac{a^{2}}{2}}\right) ^{\frac{1}{2}}\Gamma (T)U_{m}|\Psi
_{s}\rangle \noindent
\end{eqnarray}
Finally we have 
\begin{eqnarray}
&&\left( |\mathrm{exp}(0)\rangle \langle \mathrm{exp}(0)|\otimes |\tilde{\eta%
}\rangle \langle \tilde{\eta}|\otimes F_{+}\right) (V\otimes \mathbf{1}) 
\nonumber \\
&=&\left( \mathbf{1}\otimes \mathbf{1}\otimes F_{+}\right) \left( |\mathrm{%
exp}(0)\rangle \langle \mathrm{exp}(0)|\otimes |\tilde{\eta}\rangle \langle 
\tilde{\eta}|\right) V\otimes \mathbf{1}  \label{81}
\end{eqnarray}
Using (\ref{54}), (\ref{71}), (\ref{79}), (\ref{80}), and Lemma 4.3 one
easily checks the following equality. 
\begin{eqnarray}
&&W_{nm}\left( |\Psi _{s}\rangle \otimes |\tilde{\eta}\rangle \otimes |%
\mathrm{\ exp}(0)\rangle \right) \qquad (s=1,\ldots ,N)  \label{82} \\
&=&\frac{\gamma ^{2}}{N}\left( 1-e^{-\frac{a^{2}}{2}}\right) \left( |\mathrm{%
\ \ exp}(0)\rangle \otimes |\tilde{\eta}\rangle \otimes \Gamma
(T)U_{m}B_{n}^{*}\right) \otimes  \nonumber \\
&&\left( \left( 1-e^{-\frac{a^{2}}{2}}\right) |\Psi _{s}\rangle +e^{-\frac{%
a^{2}}{2}}\left( \sum\limits_{j}c_{sj}\right) \sqrt{N}\,|\Psi _{0}\rangle
\right)  \nonumber
\end{eqnarray}
For that reason we have the following Lemma

\begin{flemma}
\label{def23} For each bounded operator $A$ on $\mathcal{M}$ and $s=1,\ldots
,N$ it holds 
\begin{eqnarray*}
\vartheta _{s}(A) &:&=\left\langle W_{nm}\left( |\Psi _{s}\rangle \otimes |%
\tilde{\eta}\rangle \otimes |\mathrm{exp}(0)\right) \rangle \;,\;(\mathbf{1}%
\otimes \mathbf{1}\otimes A)W_{nm}\left( |\Psi _{s}\rangle \otimes |\tilde{%
\eta}\rangle \otimes |\mathrm{exp}(0)\rangle \right) \right\rangle \\
&=&\left( \frac{\gamma ^{2}}{N}\right) ^{2}\left( 1-e^{-\frac{a^{2}}{2}%
}\right) ^{2}\left[ \left\langle \left( 1-e^{-\frac{a^{2}}{2}}\right)
^{2}\Gamma (T)U_{m}B_{n}^{*}|\Psi _{s}\rangle \;,\;A\Gamma
(T)U_{m}B_{N}^{*}|\Psi _{s}\rangle \right\rangle \right. \\
&&+e^{-\frac{a^{2}}{2}}\left( 1-e^{-\frac{a^{2}}{2}}\right) \left(
\sum\limits_{j}c_{sj}\right) \sqrt{N}\left\langle \Gamma
(T)U_{m}B_{n}^{*}|\Psi _{s}\rangle \;,\;A\Gamma (T)U_{m}B_{n}^{*}|\Psi
_{0}\rangle \right\rangle \\
&&+e^{-\frac{a^{2}}{2}}\left( 1-e^{-\frac{a^{2}}{2}}\right) \left( \overline{
\sum\limits_{j}c_{sj}}\right) \sqrt{N}\left\langle \Gamma
(T)U_{m}B_{n}^{*}|\Psi _{0}\rangle \;,\;A\Gamma (T)U_{m}B_{n}^{*}|\Psi
_{s}\rangle \right\rangle \\
&&+e^{-a^{2}}|\sum\limits_{j}c_{sj}|^{2}N\left\langle \Gamma
(T)U_{m}B_{n}^{*}|\Psi _{0}\rangle \;,\;A\Gamma (T)U_{m}B_{n}^{*}|\Psi
_{0}\rangle \right\rangle
\end{eqnarray*}
\end{flemma}

Now from (\ref{16}) we get 
\begin{eqnarray}
&&\rho \otimes |\tilde{\eta}\rangle \langle \tilde{\eta}|\otimes |\mathrm{exp%
}(0)\rangle \langle \mathrm{exp}(0)|  \nonumber \\
&=&\sum\limits_{s=1}^{N}\lambda _{s}|\Psi _{s}\otimes \tilde{\eta}\otimes 
\mathrm{exp}(0)\rangle \langle \Psi _{s}\otimes \tilde{\eta}\otimes \mathrm{%
\ exp}(0)|  \label{83}
\end{eqnarray}
On the other hand $\left( |\Psi _{s}\otimes \tilde{\eta}\otimes \mathrm{exp}%
(0)\rangle \right) _{s=1}^{N}$ is an ONS because $\left( \Psi _{s}\right)
_{s=1}^{N}$ is an ONS. for that reason from (\ref{57},\ref{58}), (\ref{83}),
and Lemma 4.4 with $A=\mathbf{1}$ it follows 
\begin{eqnarray}
&&\tilde{p}_{nm}(\rho )=\left( \frac{\gamma ^{2}}{N}\right) ^{2}\left( 1-e^{-%
\frac{a^{2}}{2}}\right) ^{2}\left[ \left( 1-e^{-\frac{a^{2}}{2}}\right)
^{2}+Ne^{-a^{2}}\sum\limits_{s=1}^{N}\lambda
_{s}|\sum\limits_{s=1}c_{sj}|^{2}\right.  \nonumber \\
&&\left. +\sqrt{N}\,e^{-\frac{a^{2}}{2}}\left( 1-e^{-\frac{a^{2}}{2}}\right)
\sum\limits_{s=1}^{N}\lambda _{s}\sum\limits_{j=1}^{N}\left( c_{sj}\langle
\Psi _{s}\;,\;\Psi _{0}\rangle +\overline{c_{sj}\langle \Psi _{s}\;,\;\Psi
_{0}}\rangle \right) \right]  \label{84}
\end{eqnarray}
As $\left( |\mathrm{exp}(aK_{1}g_{j})-\mathrm{exp}(0)\rangle \right)
_{j=1}^{N}$ is an ONS we can calculate easily 
\[
\left\langle |\Psi _{s}\rangle \;,\;|\Psi _{0}\rangle \right\rangle =\frac{1%
}{\sqrt{N}}\sum\limits_{k=1}^{N}c_{sk} 
\]
For that reason from (\ref{84}) follows 
\begin{eqnarray}
&&\tilde{p}_{nm}(\rho )=\left( \frac{\gamma ^{2}}{N}\right) ^{2}\left( 1-e^{-%
\frac{a^{2}}{2}}\right) ^{2}\left[ \left( 1-e^{-\frac{a^{2}}{2}}\right)
^{2}+\sum\limits_{s=1}^{N}\lambda
_{s}|\sum\limits_{j=1}^{N}c_{sj}|^{2}\right.  \nonumber \\
&&\left. \left( Ne^{-a^{2}}+2\sqrt{N}\,e^{-\frac{a^{2}}{2}}\left( 1-e^{-%
\frac{a^{2}}{2}}\right) \right) \right]  \label{85}
\end{eqnarray}
Further we have $\sum\limits_{s}\lambda _{s}=1$ and 
\begin{equation}
\left| \sum\limits_{j=1}^{N}c_{sj}\right| ^{2}\le
\sum\limits_{j}\sum\limits_{k}\left| c_{sj}\right| \left| c_{sk}\right| \le
\sum\limits_{j}\sum\limits_{k}\left( \frac{|c_{sj}|^{2}}{2}+\frac{
|c_{sk}|^{2}}{2}\right) \le N  \label{86}
\end{equation}
Using (\ref{85}), (\ref{86}) and the definition of $\gamma $ (cf. (\ref{29}
)) we can estimate

\[
\begin{split}
& \left| \tilde{p}_{nm}(\rho )-\frac{1}{N^{2}}\right| \\
& =\left( \frac{\gamma ^{2}}{N}\right) ^{2}\left| \left( 1-e^{-\frac{a^{2}}{2%
}}\right) ^{2}\left[ \left( 1-e^{-\frac{a^{2}}{2}}\right)
^{2}+\sum_{s}\lambda _{s}\left| \sum_{j}c_{sj}\right| ^{2}\left(
Ne^{-a^{2}}+2\sqrt{N}e^{-\frac{a^{2}}{2}}\left( 1-e^{-\frac{a^{2}}{2}%
}\right) \right) \right] -\frac{1}{\gamma ^{4}}\right| \\
& \le \frac{1}{N^{2}}\left| \left( 1-e^{-\frac{a^{2}}{2}}\right) ^{4}+\left(
1-e^{-\frac{a^{2}}{2}}\right) ^{2}\sum\limits_{s}\lambda
_{s}|\sum\limits_{j}c_{sj}|^{2}\left( Ne^{-a^{2}}+2\sqrt{N}e^{-\frac{a^{2}}{2%
}}\left( 1-e^{-\frac{a^{2}}{2}}\right) \right) \right. \\
& \left. -\left( 1+(N-1)e^{-a^{2}}\right) ^{2}\right| \\
& \le \frac{1}{N^{2}}\left( \left| \left( 1-e^{-\frac{a^{2}}{2}}\right)
^{4}-\left( 1+(N-1)e^{-a^{2}}\right) ^{2}\right| +e^{-\frac{a^{2}}{2}
}N\left( N+2\sqrt{N}\right) \right) \\
& \le \frac{1}{N^{2}}\left( e^{-\frac{a^{2}}{2}}\left( 14+N^{2}\right) +e^{-%
\frac{a^{2}}{2}}N(N+2\sqrt{N})\right)
\end{split}
\]
That implies (\ref{41}).

\begin{flemma}
\label{def24} We use the notation $\vartheta _{s}(A)$ from Lemma 4.4. Then
for each bounded operator $A$ on $\mathcal{M}$ and $s=1,\ldots ,N$ it holds 
\begin{eqnarray*}
Z_{s}(A) &:&=\left| \frac{\vartheta _{s}(A)}{\tilde{p}_{nm}(\rho )}
-\left\langle \Gamma (T)U_{m}B_{n}^{*}|\Psi _{s}\rangle \,,\,A\Gamma
(T)U_{m}B_{n}^{*}|\Psi _{s}\rangle \right\rangle \right| \\
&\le &\frac{2e^{-\frac{a^{2}}{2}}}{\left( 1-e^{-\frac{a^{2}}{2}}\right) ^{2}}
\left( N^{2}+N\sqrt{N}+N\right)
\end{eqnarray*}
\end{flemma}

\noindent \textbf{Proof:} Using Lemma 4.4 and the estimation 
\[
\left| \left\langle \Gamma (T)U_{m}B_{n}^{*}|\Psi _{k}\rangle \;,\;A\Gamma
(T)U_{m}B_{n}^{*}|\Psi _{r}\rangle \right\rangle \right| \le \Vert A\Vert
\qquad (k,r=0,\ldots ,N) 
\]
We get 
\[
\begin{split}
Z_{s}(A)& \le \Vert A\Vert \left| \left( \frac{\gamma ^{2}}{N}\right)
^{2}\left( 1-e^{-\frac{a^{2}}{2}}\right) ^{2}\left( \tilde{p}_{nm}(\rho
)\right) ^{-1}-1\right| \\
& +\Vert A\Vert \left( \frac{\gamma ^{2}}{N}\right) ^{2}\left( 1-e^{-\frac{
a^{2}}{2}}\right) ^{4}\left( \tilde{p}_{nm}(\rho )\right) ^{-1}\left[ 2e^{-%
\frac{a^{2}}{2}}\left( 1-e^{-\frac{a^{2}}{2}}\right) \sqrt{N}\left|
\sum\limits_{j}c_{sj}\right| \right. \\
& \left. +e^{-a^{2}}N|\sum\limits_{j}c_{sj}|^{2}\right]
\end{split}
\]
Because of (\ref{85}) it follows 
\[
\begin{split}
Z& \le \Vert A\Vert \left[ \;\left| \frac{\left( 1-e^{-\frac{a^{2}}{2}
}\right) ^{2}}{\left( 1-e^{-\frac{a^{2}}{2}}\right)
^{2}+\sum\limits_{s}\lambda _{s}|\sum\limits_{j}c_{sj}|^{2}\left(
Ne^{-a^{2}}+2\sqrt{N}\,e^{-\frac{a^{2}}{2}}\left( 1-e^{-\frac{a^{2}}{2}
}\right) \right) }-1\right| \right. \\
& +\left. \frac{2e^{-\frac{a^{2}}{2}}\left( 1-e^{-\frac{a^{2}}{2}}\right) 
\sqrt{N}|\sum\limits_{j}c_{sj}|+e^{-a^{2}}N|\sum\limits_{j}c_{sj}|^{2}}{%
\left( 1-e^{-\frac{a^{2}}{2}}\right) ^{2}+\sum\limits_{s}\lambda
_{s}|\sum\limits_{j}c_{sj}|^{2}\left( Ne^{-a^{2}}+2\sqrt{N}\,e^{-\frac{a^{2}%
}{2}}\left( 1-e^{-\frac{a^{2}}{2}}\right) \right) }\right]
\end{split}
\]
Using (\ref{86}) we get 
\[
\begin{split}
& \left| \frac{\left( 1-e^{-\frac{a^{2}}{2}}\right) ^{2}}{\left( 1-e^{-\frac{%
a^{2}}{2}}\right) ^{2}+\sum\limits_{s}\lambda
_{s}|\sum\limits_{j}c_{sj}|^{2}\left( Ne^{-a^{2}}+2\sqrt{N}\,e^{-\frac{a^{2}%
}{2}}\left( 1-e^{-\frac{a^{2}}{2}}\right) \right) }-1\right| \\
& \le \frac{e^{-\frac{a^{2}}{2}}}{\left( 1-e^{-\frac{a^{2}}{2}}\right) ^{2}}
\left( N^{2}+2N\sqrt{N}\right)
\end{split}
\]
and 
\[
\begin{split}
& \frac{2e^{-\frac{a^{2}}{2}}\left( 1-e^{-\frac{a^{2}}{2}}\right) \sqrt{N}%
|\sum\limits_{j}c_{sj}|+e^{-a^{2}}N|\sum\limits_{j}c_{sj}|^{2}}{\left( 1-e^{-%
\frac{a^{2}}{2}}\right) ^{2}+\sum\limits_{s}\lambda
_{s}|\sum\limits_{j}c_{sj}|^{2}\left( Ne^{-a^{2}}+2\sqrt{N}\,e^{-\frac{a^{2}%
}{2}}\left( 1-e^{-\frac{a^{2}}{2}}\right) \right) } \\
& \le \frac{e^{-\frac{a^{2}}{2}}}{\left( 1-e^{-\frac{a^{2}}{2}}\right) ^{2}}
\left( 2N+N^{2}\right)
\end{split}
\]
That proves Lemma 4.5. $\blacksquare $\newline

We have the representation (\ref{83}) of $\rho \otimes |\tilde{\eta}\rangle
\langle \tilde{\eta}|\otimes |\mathrm{exp}(0)\rangle \langle \mathrm{exp}%
(0)| $ as a mixture of orthogonal projections. Thus from (\ref{56}) and (\ref
{57},\ref{58}) we get with the notation $\vartheta _{s}(A)$ from Lemma 4.4 
\[
\mathrm{tr}\left( \tilde{\Theta}_{nm}(\rho )A\right) =\sum\limits_{s}\lambda
_{s}\vartheta _{s}(A)\left( \tilde{p}_{nm}(\rho )\right) ^{-1} 
\]
On the other hand from Theorem 2.1 follows 
\[
\mathrm{tr}\left( \Lambda _{nm}(\rho )A\right) =\sum\limits_{s}\lambda
_{s}\left\langle \Gamma (T)U_{m}B_{n}^{*}|\Psi _{s}\rangle \;,\;A\Gamma
(T)U_{m}B_{n}^{*}|\Psi _{s}\rangle \right\rangle 
\]
Consequently we have with notation $Z_{s}(A)$ from the Lemma 4.5 
\[
|\mathrm{tr}\left( \tilde{\Theta}_{nm}(\rho )A\right) -\mathrm{tr}\left(
\Lambda _{nm}(\rho )A\right) |\le \sum\limits_{s}\lambda _{s}Z_{s}(A) 
\]
For that reason (\ref{40}) follows from Lemma 4.5, and $\sum\limits_{s}%
\lambda _{s}=1$.\newline
That completes the proof of Theorem \ref{def17}.

\end{document}